\documentclass{emulateapj}

\def\gsim{\mathrel{\rlap{\lower 4pt \hbox{\hskip 1pt $\sim$}}\raise 1pt
\hbox {$>$}}}
\def\lsim{\mathrel{\rlap{\lower 4pt \hbox{\hskip 1pt $\sim$}}\raise 1pt
\hbox {$<$}}}

\begin{document}

\title{Probing the Explosion Mechanism of Supernovae \\ 
by Radioactive Decay Gamma- and X-Rays} 

\author{
K. Maeda\altaffilmark{1,2}}

\altaffiltext{1}{Max-Planck-Institut f\"ur Astrophysik, 
Karl-Schwarzschild-Stra{\ss}e 1, 85741 Garching, Germany: 
maeda@MPA-Garching.MPG.DE}
\altaffiltext{2}{Department of Earth Science and Astronomy,
Graduate School of Arts and Science, University of Tokyo, 
3 - 8 - 1 Komaba, Meguro-ku, Tokyo
153-8902, Japan}

\begin{abstract}
Gamma- and X-rays resulting from radioactive decays provide a 
potentially powerful tool to investigate the explosion physics 
of supernovae, since the distribution and the amount of radioactive 
isotopes are strongly dependent on properties of the explosion. 
In this paper, expected features of these high energy emissions 
are presented for bipolar jet-induced explosion models, which 
are recently favored for very energetic supernovae and even for typical 
Type Ib/c supernovae. It is shown that combination of 
various observations, i.e., line-to-continuum ratio, photoelectric absorption 
cut-off energy, line profiles and luminosities, allows the unique determination 
of the explosion energy, the amount of radioactive 
$^{56}$Ni, the explosion geometry, and even the viewing orientation. 
\end{abstract}


\section{INTRODUCTION}

At the supernova (SN) explosion, various heavy elements/isotopes are produced 
as a strong shock wave passes through the inner region of the progenitor star. 
Among these newly synthesized products, important is radioactive $^{56}$Ni 
(Clayton et al. 1969). 
The decay chain $^{56}$Ni $\to$ $^{56}$Co $\to$ $^{56}$Fe 
produces $\gamma$-ray lines with average energy per decay $\sim 1.7$ MeV 
($^{56}$Ni decay with an e-folding time of 8.8 days) or 
$\sim 3.6$ MeV ($^{56}$Co decay with an e-folding time 113.7 days). 
These line $\gamma$-rays are degraded along their paths through 
the SN ejecta by 
pair production, compton scattering, and photoelectric absorption 
(e.g., Cass\'e \& Lehoucq 1994 for a review). 
The absorbed energy is the source of the optical output of a SN in 
the case of Type Ib/c SNe (SNe Ib/c) (e.g., Maeda et al. 2003 for a review). 

By looking at properties of these $\gamma$- and X-rays, it is expected 
that we can directly extract information on the distribution and the amount of 
$^{56}$Ni, which are closely related to the explosion mechanism. 
Such observations have been performed only for very nearby SN 1987A (Dotani et al. 1987; 
Sunyaev et al. 1987; Matz et al. 1988). 
Attempts to detect the high emission from distant SNe (e.g., at $\sim 10$ Mpc) have not 
been successful because of limited observational instruments in the past 
(Matz \& Share 1990; Lichti et al. 1994; Morris et al. 1997; Leising et al. 1999). 
However, now that {\it INTEGRAL} and {\it SUZAKU} have been operated, 
and some new observatories are being planned (e.g., Takahashi et al. 2001), 
it is time to revisit theoretical prediction (e.g., McCray et al. 1987; 
Woosley et al. 1987; Shibazaki \& Ebisuzaki 1988; Kumagai et al. 1989) 
taking into account recent developments in SN research, i.e., multidimensionality 
(see e.g., Maeda \& Nomoto 2003, Maeda et al. 2006ab, 
and references therein). 

Recently, multidimensional models for a supernova explosion become 
more and more popular. 
Especially, bipolar explosion models have been intensively studied 
in their explosion characteristics and nucleosynthetic features 
(e.g., Nagataki 2000; Maeda et al. 2002; Maeda \& Nomoto 2003). 
There have been some observational evidence of such 
bipolar explosions -- the 
{\it HST} image of SN II 1987A (Wang et al. 2002) 
and polarization measurements in SNe II/Ib/Ic (Wang et al. 2001; Kawabata 
et al. 2002; Leonard et al. 2002; Wang et al. 2003; Leonard et al. 2006). 

To judge a validity of a model, we have to compare model predictions 
with observations in details. For this purpose, we have been 
developing SupernovA MUltidimensional RAdIation transfer code 
{\it SAMURAI} (Maeda et al. 2006ab; Maeda 2006; 
Tanaka et al. 2006). 
The code has been applied to very energetic supernova 
(hypernova) SN Ic 1998bw associated with a Gamma-Ray Burst. 
By comparing the model predictions with optical spectra 
and light curve of SN 1998bw in details, 
Maeda et al. (2006ab) concluded that SN 1998bw is the bipolar explosion 
viewed from the jet direction. 

In this paper, we present the model prediction for radioactive decay 
$\gamma$- and X-rays from supernovae driven by 
bipolar jets. Details are presented 
in Maeda (2006) (see also Hungerford et al. 2003, 2005). 
Although the model has been applied to hypernovae, 
we believe that most of the model predictions apply even to usual 
SNe Ib/c which also show some evidence of asphericity 
(e.g., Wang et al. 2001). 

\begin{figure*}[t]
\begin{center}
\hspace{-4cm}
       \begin{center}
	\begin{minipage}{0.47\textwidth}
		\epsscale{1.0}
		\plotone{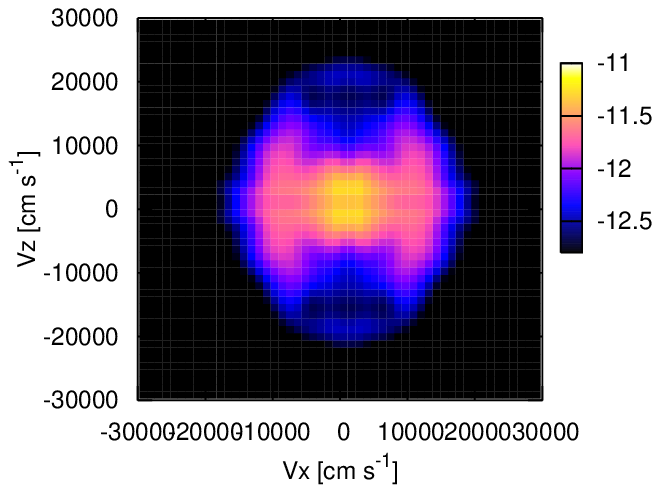}
	\end{minipage}
       \begin{minipage}{0.47\textwidth}
		\epsscale{1.0}
		\plotone{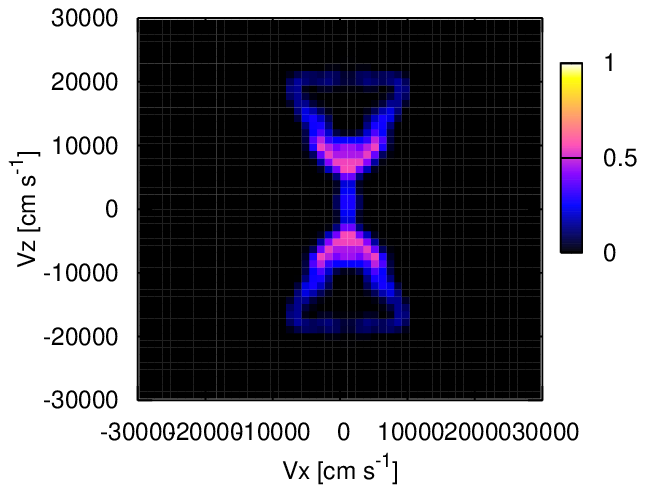}
	\end{minipage}
	\end{center}
\end{center}
\caption[]{Model A (Maeda et al. 2002) with $E_{51} = E/10^{51}$ erg = 20. 
{\it Left:} Density distribution (g cm$^{-3}$) on a logarithmic scale 
at $t = 10$ days. The ejecta are already freely expanding (so called 
homologous expansion or Hubble flow), so the distribution is shown in 
velocity space. {\it Right:} Mass fraction of $^{56}$Ni on a liner scale.   
\label{fig1}}
\end{figure*}

\section{METHOD AND MODELS}
We use a fully time- and energy-dependent radiation transfer scheme 
to follow the high energy photons resulting from 
$^{56}$Ni/Co/Fe decay chain. This is done by using a Monte-Carlo method. 
Line $\gamma$-rays are produced according to the decay probabilities, 
then the photons are followed along their paths in which they suffer from 
pair creation, compton scattering, and photoelectric absorption in 
expanding 3D SN ejecta.  
The detail is described in Maeda (2006). 

We adopt the bipolar model A 
from Maeda et al. (2006b), which successfully 
reproduces the optical observations of hypernova SN 1998bw. 
The hydrodynamic and nucleosynthetic features have been examined 
in detail in Maeda et al. (2002).  
For comparison, we also examine the spherical model S. 
The ejecta structure (density and $^{56}$Ni) is shown in Figure 1. 
In Model A, $^{56}$Ni is distributed preferentially along the 
jet axis because the materials in this direction experience 
high temperature (i.e., $\gsim 5 \times 10^9$ K) at the explosion. 
On the other hand, the density is enhanced in the equatorial plane 
(showing a disk-like structure), since only a weak shock wave 
pushes materials outward in the equatorial direction 
(Maeda et al. 2002; Maeda \& Nomoto 2003).

\section{RESULTS}

\subsection{Spectra}

Figure 2 shows the high energy spectra at $t = 25$ and $50$ days, 
which are obtained by our simulations. 
Hereafter, $t$ is the time after the explosion. 

The aspherical model A has three unique features as compared 
to the spherical model S. 
\begin{itemize}
\item[(1)] {\bf Large line-to-continuum ratio: } 
Model A shows a large line-to-continuum ratio, especially 
early on. 
In Model A, 
the source of the high energy emission, i.e., $^{56}$Ni/Co, is abundantly present 
around the jet axis ($z$-direction). 
In this direction, the density is small, thus the optical depth is small, 
as the consequence of the jet-induced explosion. 
Thus, few photons suffer from multiple 
compton scatterings which create the continuum in Model A. 
Since the situation does not 
change very much as time goes by, the line-to-continuum ratio 
does not show noticeable temporal evolution. 

Also, the line-to-continuum ratio is not sensitive to the viewing orientation. 
Both observers in the $z$- and $r$-direction look at the same region, i.e., 
the $^{56}$Ni/Co-rich jet. The line-to-continuum ratio is determined by 
local physical conditions in the emitting region, which is same irrespective 
of the observer's direction. Thus, it is nearly independent from 
the observer's direction. An exception is a line profile, which 
is discussed in \S 3.2. 

In Model S, a situation is different. 
The $^{56}$Ni/Co-rich region is surrounded by massive envelopes in Model S.  
The line photons experience many scatterings, creating the high level of the 
continuum. In Model S, the envelope becomes less and less optical thick 
due to the SN expansion, thus the continuum level becomes lower and lower as time goes by. 

\item[(2)] {\bf Large cut-off energy: }
The hard X-ray cut-off, which is created by photoelectric absorption, 
is at higher energy in Model A than Model S. 
In Model A, the last scattering point is close to the $^{56}$Ni-enhanced jet, 
thus the photoelectric absorption takes place in the Fe-rich materials. 
Since the situation does not change as time goes by, the cut-off energy 
does not show temporal evolution. 
This behavior is again insensitive to the observer's direction. 

On the other hand, Model S shows temporal evolution. 
Early on, the last scattering takes place near the surface where O is the dominant 
element. Later on, the last scattering point moves deeper to the inner region 
where Fe is the dominant element. Thus, the cut-off energy moves from lower to 
higher energy. 

\item[(3)] {\bf Early emergence: }
In Model A, escape of the high energy photons occurs earlier than in Model S 
if the kinetic energy of the expanding SN ejecta is the same. 
This is expected since $^{56}$Ni is produced along the axis where 
the optical depth is low in Model A. 

\end{itemize}

\begin{figure*}[t]
\begin{center}
\hspace{-4cm}
       \begin{center}
	\begin{minipage}{0.47\textwidth}
		\epsscale{1.0}
		\plotone{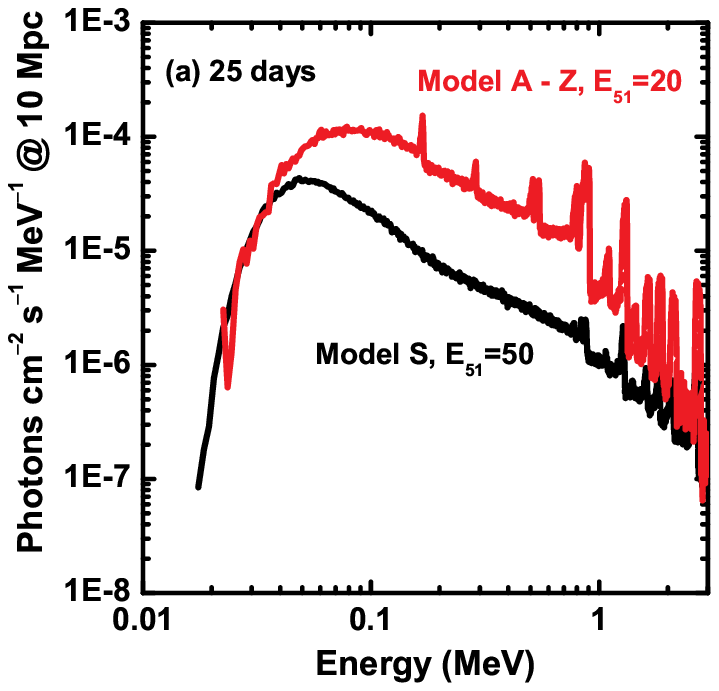}
	\end{minipage}
       \begin{minipage}{0.47\textwidth}
		\epsscale{1.0}
		\plotone{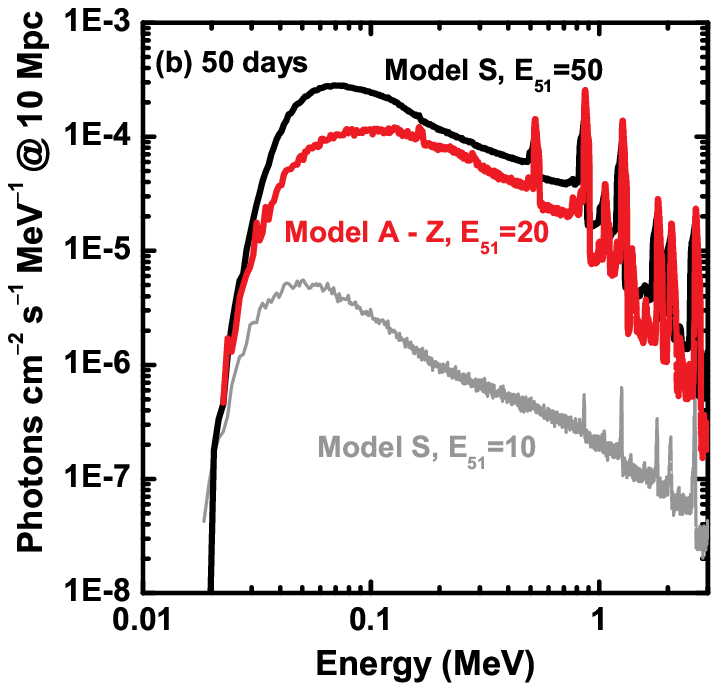}
	\end{minipage}
	\end{center}
\end{center}
\caption[]{Synthetic hard X- and $\gamma$-ray spectra (a) at $t = 25$ days 
and at (b) $t = 50$ days. 
\label{fig2}}
\end{figure*}

\begin{figure*}[t]
\begin{center}
\hspace{-4cm}
       \begin{center}
	\begin{minipage}{0.47\textwidth}
		\epsscale{1.0}
		\plotone{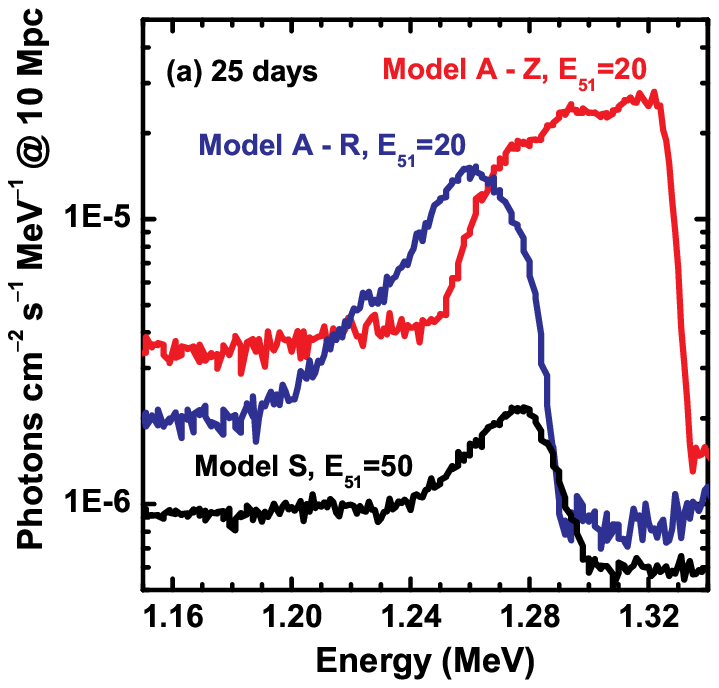}
	\end{minipage}
       \begin{minipage}{0.47\textwidth}
		\epsscale{1.0}
		\plotone{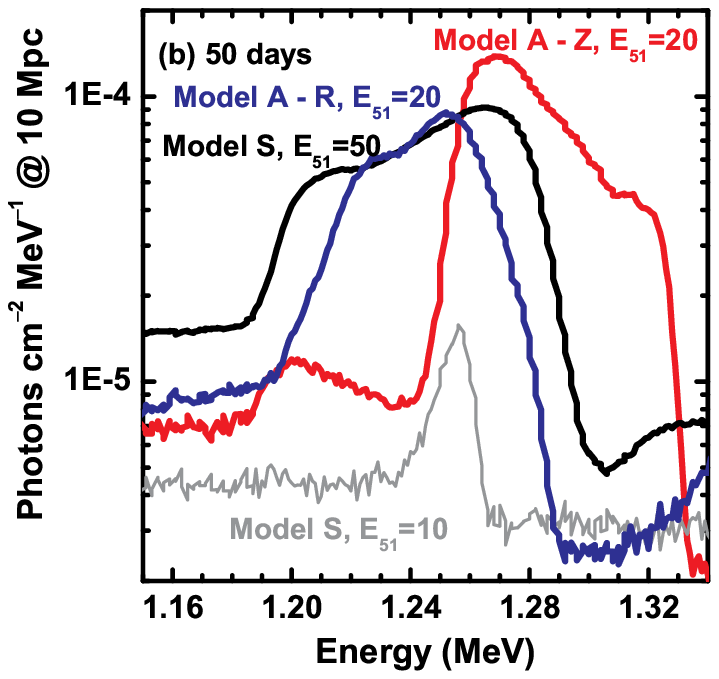}
	\end{minipage}
	\end{center}
\end{center}
\caption[]{Profiles of the synthetic $^{56}$Co 1,238 keV line (a) at $t = 25$ days 
and (b) at $t = 50$ days. 
\label{fig3}}
\end{figure*}

\subsection{Line Profiles}

Line profiles are also different for models A and S. 
As an example, Figure 3 shows the line profiles of $^{56}$Co 1,238 keV line. 
Other lines have similar properties, but the line blending should be taken into account 
(e.g., $^{56}$Ni 812 keV and $^{56}$Co 847 keV). 

Model A yields a blueshifted line if viewed from the 
jet ($z$-) direction. In this case, an observer views the 
optically thin $^{56}$Ni/Co-rich region from the jet ($z$-) direction, and 
the emission from the other hemisphere is blocked by the optically thick central region. 
The central region continues to be optically thick at $t \lsim 100 - 200$ days, 
thus the line keeps to be blueshifted. 

If Model A is viewed from the $r$-direction, then two $^{56}$Ni/Co-rich jets 
in both hemispheres can be seen. In this case, the emitting regions (the jets) 
are moving to the directions perpendicular to the observer. 
Thus, the line is centered at the rest wavelength when viewed from the $r$-direction. 
The situation does not change as time goes by, thus the line center does not evolve 
as a function of time in Model A (irrespective of the observer's direction). 

However, the situation is different for Model S. 
Early on, only the outer region of the $^{56}$Ni/Co-rich region is seen, 
yielding the blueshifted line. As time goes by, the observer can look into 
the deeper region (because of the ejecta expansion and the density decrease), 
thus the line center continuously moves from the blue to the rest wavelength.

\section{CONCLUSIONS AND DISCUSSIONS}

\begin{deluxetable}{cccc}[t]
 \tabletypesize{\scriptsize}
 \tablecaption{Characteristics
 \label{tab:summary}}
 \tablewidth{0pt}
 \tablehead{
   \colhead{Features}
 & \colhead{Model S}
 & \colhead{Model A - Z}
 & \colhead{Model A - R}
}
\startdata
Line-to-Cont\tablenotemark{a}           & Small & Large & Large\\
Cut-off E\tablenotemark{b}              & Low   & High  & High\\
Evolution of spectra\tablenotemark{c}   & Large & Small & Small\\
Line profile\tablenotemark{d}           & Evolved & Blue & Rest\\
Evolution of Lines\tablenotemark{e}     & Blue to Rest & Small & Small
\enddata
\tablenotetext{a}{The line-to-continuum ratio, determined by compton scatterings.}
\tablenotetext{b}{The cut-off energy, determined by photoelectric absorptions.}
\tablenotetext{c}{The evolution of the Line-to-continuum ratio and the cut-off energy.}
\tablenotetext{d}{The line profile of isolated lines (especially at 1,238 keV).}
\tablenotetext{e}{Temporal evolution of the line profile.}
\end{deluxetable}

We have shown in this paper that expected observed features 
in the high energy emission are highly dependent on the geometry of the explosion. 
Models A and S have quite different characteristics, thus 
these model can be distinguished by observations of the high energy emission. 
Among the features, the line profile is very sensitive to the 
viewing orientation in Model A. Thus, if we can perform deep observations 
to detect the line profile, then even the viewing orientation can be distinguished. 
In addition, temporal evolution is very different for models A and S. 
The expected features are summarized in Table 1. 

For the sensitivity of {\it INTEGRAL} ($3 \sigma$ in $10^{6}$ sec), 
we estimate that the maximum distances within which the $\gamma$-rays 
are detectable are $\sim 500$ kpc, $4$ Mpc, and $7$ Mpc for 
SNe II, canonical SNe Ib/c, and hypernovae, respectively. 
Since SN II 1987A, the nearest SNe for each type are 
SN II 2004dj ($\sim 3$ Mps), SN Ic 1994I ($\sim 7$ Mpc), and 
the hypernova SN Ic 2002ap ($\sim 10$ Mpc). 
Thus, the sensitivity of {\it INTEGRAL} is unfortunately not 
enough to detect these core-collapse SNe in the pace of 1 per 2 decades, 
unless very nearby SNe like SN 1987A fortunately appears in the sky. 

If new generation hard X- and $\gamma$-ray telescopes are designed 
to achieve a sensitivity $\sim 1$ magnitude or two better than the current 
instrument (Takahashi et al. 2001), 
then it is expected that the high energy emission from core-collapse 
SNe can be detected $\sim 1$ SN per year. 
This sensitivity will lead to a comprehensive study of the SN explosive 
physics through observing the high energy emission.

\acknowledgements

The author is the JSPS Research Fellow.

%\onecolumn

\end{document}